\documentclass[aps,prl,preprint,superscriptaddress,showpacs]{revtex4}

\usepackage{amssymb}
\usepackage{epsfig}
\usepackage{amsmath}
\usepackage{times}

\begin{document}

\title{Exploring network structures in feature space}

\author{Xiaofeng Gong}
\affiliation{Temasek Laboratories, National University of Singapore, Singapore 117508}
\affiliation{Beijing-Hong Kong-Singapore
Joint Center of Nonlinear and Complex systems (Singapore), National University of Singapore, 
Singapore 117508}
\author{Shuguang Guan}
\affiliation{Temasek Laboratories, National University of Singapore, Singapore 117508}
\affiliation{Beijing-Hong Kong-Singapore
Joint Center of Nonlinear and Complex systems (Singapore), National University of Singapore, 
Singapore 117508}

\author{C.-H. Lai}
\affiliation{Department of Physics, National University of Singapore, Singapore 117542}
\affiliation{Beijing-Hong Kong-Singapore
Joint Center of Nonlinear and Complex systems (Singapore), National University of Singapore,
Singapore 117508}

\begin{abstract}
We propose a multi-phase approach to explore network structures. In this method, structure analysis is not carried out on the observed network directly. Instead, certain similarity measures of the nodes are derived from the network firstly, which are then projected onto an appropriate lower-dimensional feature space. The clustering structure can be defined in the feature space, and analyzed by conventional clustering algorithms. The classified data are finally mapped back to the original network space if necessary to complete the analysis of network structures. By mapping onto the feature space, some difficulties due to the diversity of micro-structures and scale of the network can be circumvented.  This makes it possible for the proposed method to deal with more general structures such as detecting groups in a random background, as well as identifying usual community structures in networks.
\end{abstract}

\maketitle

Networks (or graphs) are natural representations for many complex systems, where the vertices (or nodes) stand for certain entities, and the edges (or links) represent the inter-connections (dynamical or stationary) which can be physically existing channels, or certain relationship in a more general sense. There are various substructures in complex networks in general. 
When the underlying system is well understood, we are usually able to figure out different substructures in light of the global picture of the whole system. Sometimes, even very detailed structure such as a single edge can be identified and related to certain function. Inversely, it is interesting to think about whether deeper insights of the underlying system (such as unseen relationships) can be inferred by investigating the strucure(s) of a representive network.  In biological networks for example, it is widely believed that the modular structures play a crucial role in biological functions\cite{Hartwell1999,Ravasz2002}. 
Unfortunately, when inferring from the functions, the network links appear bewildering, and the intrinsic structures of the network is often obscured, not to mention their relation to the functions of the underlying system. In many situations, identification of communities is a highly nontrial problem.  

Currently, there is no universally accepted rigorous definition for communities in a network. It is usually thought of as subsets of nodes which are densely interconnected (intra-cluster) and sparsely connected to the rest of the network (inter-cluster). Based on this intuitive understanding, many methods are proposed to detect and identify communities in networks\cite{NewmanGirvan2004,Newman2004,Newman2006,Duch2005,Wu2004,Reichardt2004}. 

However, there are some important aspects which are largely ignored. Firstly, most of the studies focus on networks which are exclusively covered by communities. In other words, each node has to be assigned to one community or another. This is not the case in many realistic situations. It is quite possible that an otherwise sparsely connected network has one or several groups of nodes densely interconnected. Although the nodes in these groups can be regarded as in clusters, a conceptual difficulty would arise if the rest of the nodes had to be assigned to one or more clusters, since there is obviously not much difference between the intra- and inter-cluster connection densities for these nodes.  In such a case, the whole picture is more like one where there are some substructures embedded in a certain background. Detecting and identifying these small communities is certainly very useful in practice.

Another consideration is that a structure is essentially a relative concept. Inter connections within any subset of nodes, by themselves, say nothing about whether if these nodes can be identified as a community.  For example, even fully connected group of nodes does not form a community if each of them connects to all outside nodes, while several sparsely connected nodes can be a legimate community if they effectively do not link to outside peers. One of the consequences is that prominent network structures may depend on the scale of the investigation. For instance, consider a network with a multi-centered structure, where all peripheral nodes are connected to several mutually connected center nodes. Such a structure, if it exists in a large sparse network, can be considered as one community. However, if the investigation scale is zoomed in to focus on this structure, it is more reasonable to take the center nodes only as a community. For an extreme example, let us consider a bipartite subnetwork. Again, in a large network, this subnetwork can be identified as a community in the usual sense. If the whole network has an approximately bipartite structure, one cannot define a community strucutre in the usual sense, even though there are two families of nodes with clearly distinct connection patterns. A good analysis method should be able to adapt with the network scale automatically.

In this paper, we propose a different approach to analyze network structures, which allows us to avoid these difficulties. In this method, the structure analysis is not performed on the network data directly. Instead, the network is first projected onto some appropriate low-dimensional feature space based on some similarity measures of the nodes. In the feature space, the mapped data points corresponding to nodes with similar characteristics always group together and form certain structures with different densities. Cluster strucutures thus can be easier defined in the feature space based on various criteria as in conventional clustering analysis\cite{Jacob2007,Fukunaga1990}, and can be identified by well understood clustering algorithms such as the K-means method. 

One of advantages to carry out clustering analysis in feature space is that the structure appearing in feature space simply depends on the relative similarity measures of the nodes. Data points in a well defined cluster in feature space may not always correspond to a community in the original network in the sense discussed above. For example, the nodes which contribute the background of random connections may form a clear cluster in feature space; and two families of nodes consisting a bipartite structure may appear as two clusters in the feature space. By this way, the intricacies arising from the structure heterogeneity of the network can be circumvented. In practice, an extra step of mapping the clusters in feature space back to the original network may be taken, depending on the problem at hand, to further investigate their implications in the context of the original network, e.g., if two clusters in feature space actually make up a bipartite structure and need to be merged.
 
The proposed analysis method thus works at four different phases: 1) we need to derive similarity measures of nodes of the network under investigating; 2) we extract relevant features from the similarity measures, and project them onto an appropriate lower-dimensional feature space; 3) we carry out conventional clustering analysis in feature space; and 4) we interpret the analysis results in the context of the original network.  In the remaining parts of the paper, we will first describe a specific implementation of the algorithm, and the method is then applied to some model networks to demonstrate its advantages.

Given an undirected network, there are many ways to measure the similarity between nodes. For the purpose of structure analysis, the most straightforward one is based on connection patterns of the nodes, which are completely encoded by the corresponding rows of the adjacency matrix $A_{N \times N}$ associated with the network. Let $\{d_j, j=1,2,\cdots,N\}$ be the degrees of the nodes. The column vector $s_j = a_j -D/N$ can be regarded as the centered connection pattern of node $j$, where $a_j$ is $j$th column of $A$ and $D=[d_1,d_2,\cdots,d_N]^T$. The internal correlated structure of $s_j$ can be studied by principal component analysis (PCA). PCA is mathematically defined\cite{Jolliffe2002} as an orthogonal linear transformation that transforms the data to a new coordinate system such that the greatest variance by any projection of the data comes to lie on the first coordinate (called the first principal component), the second greatest variance on the second coordinate, and so on. PCA is theoretically the optimum transform for given data in least square terms. The first few principal components thus can be taken as effecitve features of the original data that contribute most to its variance. We prefer to extract these features through singular decomposition of the data matrix $S=[s_1,s_2,\cdots,s_N]$, which can be applied even when only partial information is available.

Let $S=U \Sigma V^T$ be the singular decomposition of $S$, where $U_{N\times N}$ and $V_{N\times N}$ are the left and right singular vectors respectively, and $\Sigma_{N\times N}$ is a diagonal matrix whose elements are singular values. The connection patterns are then projected onto the feature space which is spanned by a few leading singular vectors $u_k,k=1,2,\cdots,M$. The mapped data points $F=S^T [u_1,u_2,\cdots,u_M]$ in feature space will be further analyzed. The dimension of the feature space used depends on the problem on hand. Usually, a low dimensional feature space (e.g. $2D$ or $3D$) is preferred. This is not only because of the lower computational load for later clustering analysis: in a low dimensional feature space, a clear picture of the distribution of mapped data often provides good suggestion of crucial parameters such as cluster number and initial partitions.  The singular value spectrum gives useful information regarding the dimension of the feature space. In general, when cluster structures are clear, several leading singular values are significantly above the rest and suggest the proper number of singular vectors involved in the features set. However, the gap will be smeared as the cluster structure becomes vague. 

After projecting the network onto an appropriate feature space, clustering anaysis can be carried out on those mapped data.  For this, many well developed algorithms are ready to be used. In this study, we apply an improved version of the K-means algorithm\cite{Jacob2007}, which is one of the most widely used and well understood clustering methods. K-means is an iterative algorithm to minimize the objective function $Q$, which is the sum of point-to-centroid distances, summed over all $K$ clusters.  Starting from the initial assignment of the data points to each cluster and determining the corresponding cluster centroid (Euclidean distance in feature space are used), each iteration consists of reassigning points to their nearest cluster centroid, all at once, followed by recalculation of cluster centroids.  This procedure will be stopped if no improvement could be achieved. 

To improve the overall performance, a refinement phase is applied. After a stable partition $\Pi=\{\pi_1,\pi_2,\cdots,\pi_K\}$ has been created by the above procedure, the first variation partition $\Pi'$, which can be obtained by removing a single point from a cluster $\pi_i$ and assigning this point to an existing cluster $\pi_j$, is generated. The cluster centroids are then recalculated. If a smaller $Q$ or a better partition can be found in any $\Pi'$, the ordinary K-means procedure in the first step is restarted again from this new partition. The two procedures above are repeated until no improvement can be achieved.  The algorithm can still converge to a local optimum, even using the refinement step, which in this case is a partition of points in which moving any single point to a different cluster increases the total sum of distances.  This problem can only be solved by a clever (or lucky, or exhaustive) choice of starting points. In our simulation, the same procedure is repeated $10$ times using random initial conditions, and the best one is picked as the final result.  

To illustrate how the multi-phase algorithm works, we first apply it to the modular network studied in \cite{NewmanGirvan2004}, which is a random network consists $128$ nodes with $4$ densely connected clusters (each contains $32$ nodes).  The connection patterns in this network can be modeled by $2$ parameters concisely: $p_{in}$ and $p_{out}$, where $p_{in}$ stands for the connection probability of two nodes in the same cluster, and $p_{out}$ the connection probability of two nodes in different clusters. The values of $p_{in}$ and $p_{out}$ are chosen to make the expected degree of each node equals to $16$. In this model, every node in each community has the same connection pattern statistically, and the communities cover the whole network.

\begin{figure}[thb]
   	\begin{center}
	\epsfig{figure=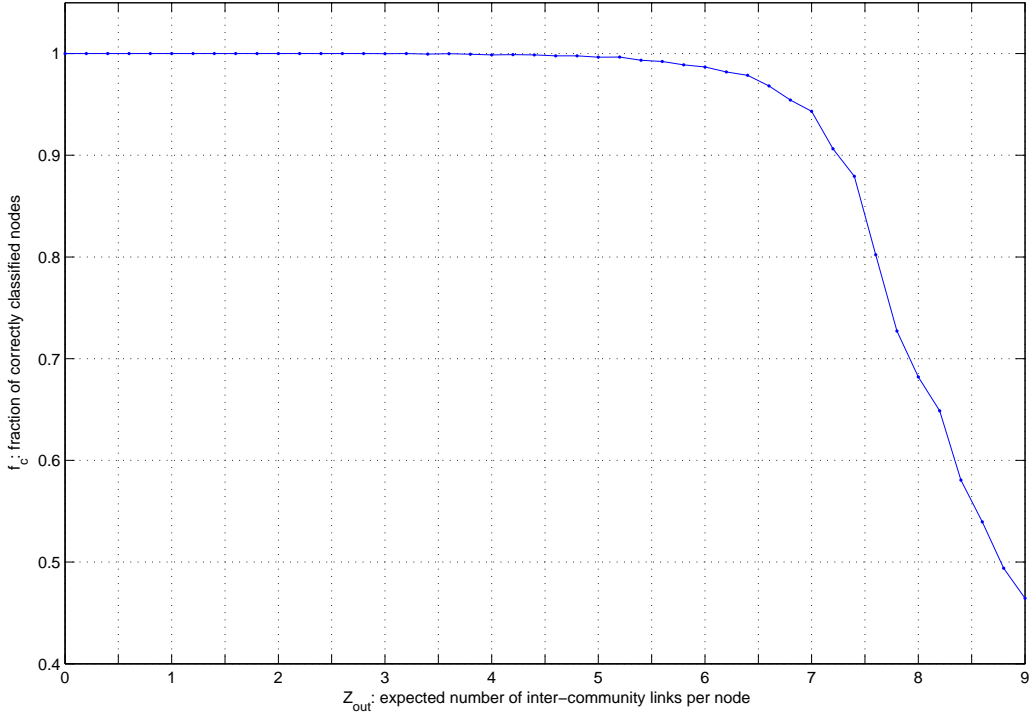,width=1\linewidth}
	\end{center}
     \caption{Clustering performance of the algorithm applying on the modular network as described in text. The results are the average of $50$ different realizations based on the same model.} 
	\label{fig:fig1}
\end{figure}

\begin{figure}[thb]
   	\begin{center}
	\epsfig{figure=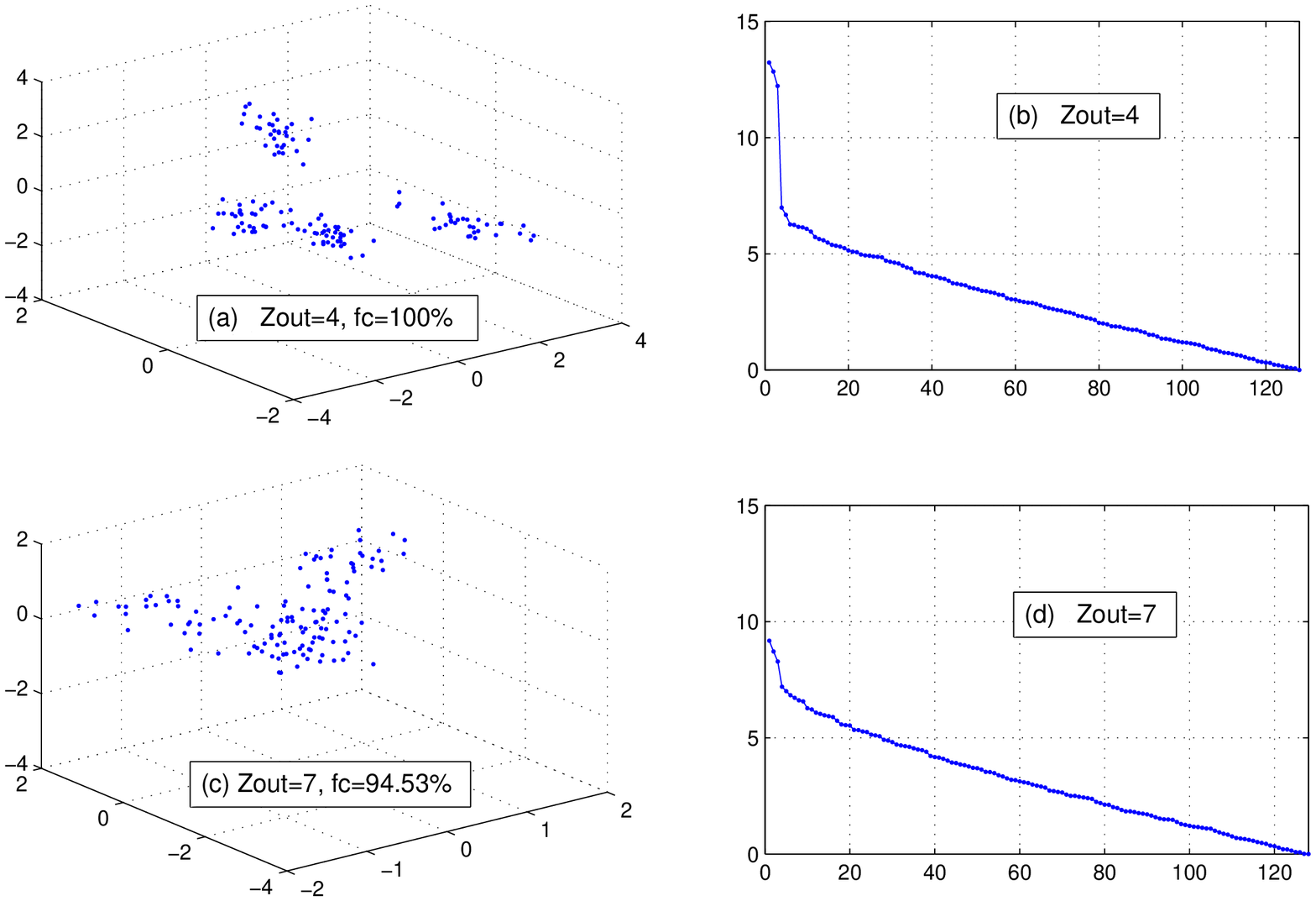,width=1\linewidth}
	\end{center}
     \caption{Distributions of projected connection patterns of nodes in $3D$ feature space for two particular cases. The network model is the same as in figure 1. The number of inter-community edges are indicated in all figures. The clustering performance (fraction of correctly classified nodes) are indicated in the figure (a) and (c).} 
	\label{fig:fig2}
\end{figure}

The overall performance of clustering on this network is shown in figure \ref{fig:fig1}, where the fraction of nodes classified correctly is shown as a function of the mean number of inter-cluster links $z_{out}$. The results are the average of $50$ different realization of random networks based on the same model. It can be seen clearly in the figure that the clustering results are almost perfect when $z_{out}$ is relative small until $z_{out}$ approachs $6$. The errors of misclassification increase quickly after $z_{out} \approx 7$. However, in contrast to the results shown in \cite{NewmanGirvan2004}, in this study, even when $z_{out}$ is around $8$, the error is still significantly smaller. In fact, at this point, though the intra-cluster links $z_{in}$ is the same as the inter-cluster links $z_{out}$, $p_{in}$ is still significantly larger than $p_{out}$, since there are much more outside nodes than inside nodes ($3$ times). 

To better reveal how the method works, more details are shown in figure \ref{fig:fig2}. The clustering structures can be seen clearly from the projections of the orignal connection patterns in $3D$ feature space (as shown in figure 2(a)) when $z_{out}=4$. In this case, first $3$ leading singular values gap up significantly (as shown in figure 2(b)), suggesting the appropriate feature space dimension. However, this kind of information become less useful when modular structures become vague (see figure 2(c) and figure 2(d) in the case when $z_{out}=7$ )

When applying the method to some real example such as the Karate club network and the dolphins social network, we first determine the dimension of the feature space by observing the singular spectrum. The cluster number then have to be guessed based on the distribution of the data points in the feature sapce. The clustering results are similar to what reported in the literatures\cite{NewmanGirvan2004,Newman2004}.

Now let us consider more interesting examples to address the points we discussed earlier. Firstly, consider a random network with $N$ nodes where a relatively densely connnected group of nodes are embedded. The purpose of the analysis is to identify this group. To construct a network having a desired structure, we need a random netwrok with a preassigned degree distribution, which is generated by the following procedure.
\begin{itemize}
\item Use a configuration model to generate a random network with the required degree distribution\cite{Newman2001}. In this network, multiple and self-connections are allowed, and will be removed in the next steps.
\item To remove each self-connections of node $k$, we first find two connected nodes $i$ and $j$ which are not connected to node $k$. A pair of new edges $ki$(and $ik$) and $kj$(and $jk$) are created, while one original self-connection of node $k$ and the edge $ij$(and $ji$) are removed. 
\item To remove one of the multiple connections between node $i$ and node $j$, we first search for a pair of connnected nodes $i'$ and $j'$. Each of them does not connected to nodes $i$ and $j$ simultaneously. One of the multiple connections $ij$ (and $ji$) and the edges $i'j'$ (and $j'i'$)  are replaced by the edges $ii'$ (and $i'i$) and $jj'$ (and $j'j$). 
\item The rewiring procedure is repeated until there is no multiple and self-connection. If no legitimate nodes can be found to be rewired to, a random network candidate is regenerated using the configuration model. 
\end{itemize}
The smaller hidden group is modeled by a random network of $M$ nodes with average degree $\langle n\rangle$. These two networks are then superimposed randomly and repeated edges are removed. 

By applying the proposed multi-phase clustering analysis, we get two sets of nodes finally. The smaller set $S$ is taken to be an estimate of the hidden group, and the other set $B$ corresponds to the background nodes. Both $S$ and $B$ may contain nodes coming from the group and the background in general. A complete measurement of clustering performance thus requires a $2 \times 2$ confusion matrix\cite{Fukunaga1990}. Since we mainly focus on the identification of the hidden group, only two quantities corresponding to two terms in the confusion matrix are used to measure the performance. Suppose the sizes of $S$ and $B$ are $n_s$ and $n_b$; we have $N = n_s +n_b$. Let $n_s = n_t +n_f$, where $n_t$ and $n_f$ are the numbers of nodes in $S$ coming from the hidden group and background respectively. Then the quantity $q_1= n_t/M$ measures the fraction of nodes in the hidden group which are correctly assigned to $S$. To further describe the quality of $S$, the quantity $q_2 = n_f/n_s$ is used to measure the fraction of misclassified nodes in $S$. $q_1$ and $q_2$ together give the overall performance of a particular result. For a perfect partition, we have $q_1=1$ and $q_2=0$. A good clustering result shall show large $q_1$ and small $q_2$ simultaneously. In practice, usually one is treated as a more important measure than the other depending on the nature of the problems analyzed. 

\begin{figure}[thb]
   	\begin{center}
	\epsfig{figure=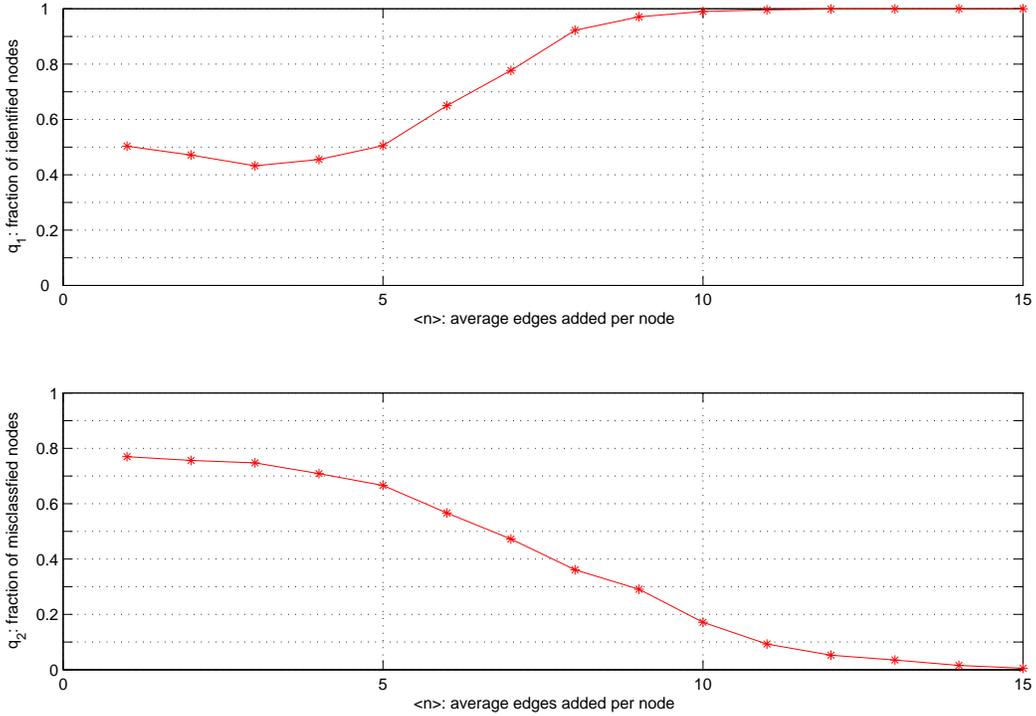,width=1\linewidth}
	\end{center}
     \caption{Clustering performance of the algorithm as the function of the average degree of the nodes in the hidden group. The results are averaged over $100$ different realizations.} 
	\label{fig:fig3}
\end{figure}

In figure \ref{fig:fig3}, we show the averaged results of clustering analysis on $100$ different realizations of random networks described above. The performance depends on the connection density in the hidden group. The results are acceptable even when the average number of edges in the group is similar to that of the background. For instance, in one particular test, the degree of the background network is uniformly distributed in the range of $[3,21]$, and the average degree of the whole network ($N=200$) is $12$. 
%
%
We then construct a small network ($M=40$) with average degree $\langle n\rangle$=8.  After superimposing randomly these two networks, the average number of inside edges for each node in the group is slightly less than $10.4$, while the average number of outside edges for each node is about $9.6$. In this case, $q_1$ is above $0.9$ and $q_2$ is around $0.35$. The value of $q_2$ is a bit larger than expected due to the fact that by chance there are some nodes in the background which show very similar connection patterns as the node in the hidden group and cannot be classified correctly by the algorithm. 

\begin{figure}[tbp]
   	\begin{center}
	\epsfig{figure=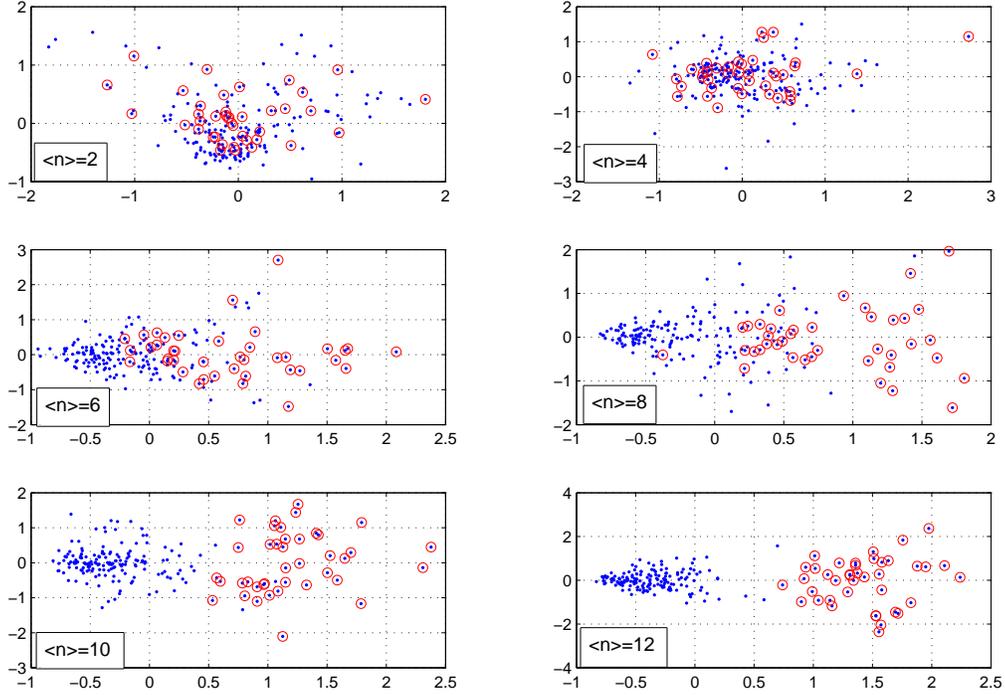,width=1\linewidth}
	\end{center}
     \caption{Distributions of mapped connection patterns in $2D$ feature space for several particular cases. The average degree in the hidden group $\langle n\rangle$, as indicated in each sub-figures, changes from from a low value for which the group structure is not able to be recognized completely to a high value for which the clusters are clearly visible and can be idientified correctly. In these figures, the (blue) dots represents the nodes of the whole network, and those with (red) circles correspond to the group members.} 
	\label{fig:fig4}
\end{figure}

Typical distributions of mapped data points in $2D$ feature space are shown in figure \ref{fig:fig4} when the connection density within the hidden group changes. It is intresting to observe how the points corresponding to nodes in the background group together. This demonstrates the advantage of making clustering analysis in feature space. 

\begin{figure}[tbp]
   	\begin{center}
	\epsfig{figure=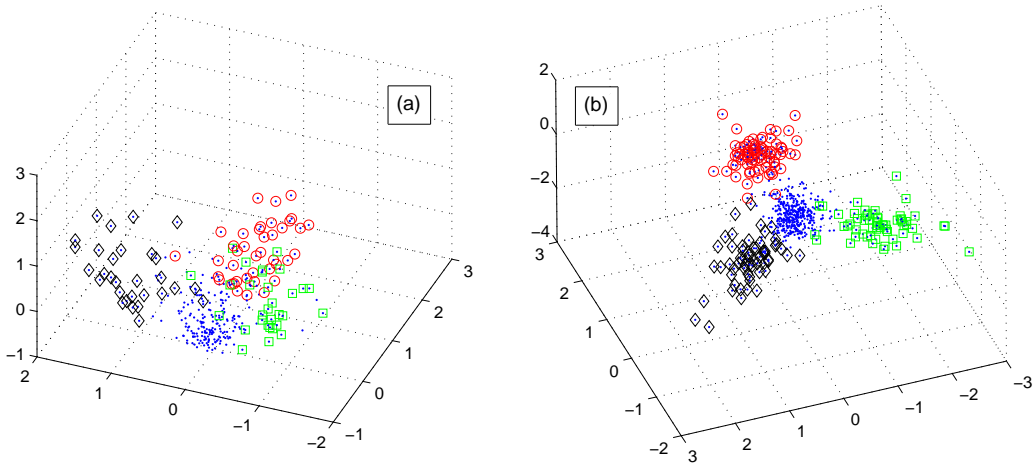,width=1\linewidth}
	\end{center}
     \caption{Distributions of the mapped connection patterns of nodes in a $3D$ feature space for two particular cases. The whole network consists of one uniform cluster, one approximate bipartite cluster and the background of a random network as described in text. In the figures, the (blue) dots is corresponding to all nodes in the whole network, where those with (red) circles represent the nodes in uniform cluster, and those with (green) squares and those with (black) dimonds represent the nodes in two different families of bipartite cluster.  In case (a), the average numbers of inner edges to form unifom and bipartite cluster are $\langle n_1\rangle=10$ and $\langle n_2\rangle=7$ respectively. In case (b), the numbers are $\langle n_1\rangle=10$ and $\langle n_2\rangle=9$. In the bipartite cluster, the average degrees of all nodes are the same. The construction of the background network is the same as that in figure 3 and figure 4, except for different size.} 
	\label{fig:fig5}
\end{figure}

We also study a more complicated situation, where on the backgroud of a random network ($N=300$), there are two clusters with different micro-structures. One of them is a uniformly densely connected cluster ($M_1=40$) and the other is an approximately bipartite cluster ($M_2=M_{2a}+M_{2b}=30+30$). The network is constructed by superimposing randomly a densely connected subnetwork and a perfect bipartite subnetwork on the background of a random network (in a similar way as in the case of figure 3 and 4). In a perfect bipartite network, all nodes can be divided into two families. Any node in one family can only be connected to the nodes in the other family. In our example, two nodes in the same family may be connected due to the existing connection in background network. In figure \ref{fig:fig5}, the projections of the original connection patterns in a $3D$ feature space are shown for two particular cases. $4$ clusters can be identified satisfactorily by the method, where one corresponding to the dense uniform cluster, two of them corresponding to two different families of the bipartite cluster, and the last one for background nodes (as indicated in the figure). As shown in the case of figure 5(b), more accurate results can be obtained when there are more connections in the  bipartite cluster. The clustering performances of both cases are described by confusion matrices as shown in Table I. 

\begin{table}[hbt]
	\begin{tabular}{cc|c|c|c|}
                    	        &\multicolumn{4}{c}{Assigned classes} \\ \cline{2-5}
   				 &     	   &uniform 	&bipartite  	&background 	\\ \cline{2-5}
	  Actual\  \	 &uniform	   &40		&0			&0			\\ \cline{2-5}
	  classes\ \ 	 &bipartite  &1		&55			&4			\\ \cline{2-5}
				 &background &4		&21			&175			\\ \cline{2-5}
					&\multicolumn{4}{c}{(a)}
    
	\end{tabular}
	\begin{tabular}{cc|c|c|c|}
                    	        &\multicolumn{4}{c}{ Assigned classes} \\ \cline{2-5}
   				 &     	   &uniform 	&bipartite  	&background 	\\ \cline{2-5}
	Actual\  \	 &uniform	   &40		&0			&0			\\ \cline{2-5}
	classes\ \ 	 &bipartite  &0		&57			&3			\\ \cline{2-5}
				 &background &5		&1			&194			\\ \cline{2-5}
					&\multicolumn{4}{c}{(b)}
	
	\end{tabular}
\caption{Confusion matrix for clustering analysis in cases of figure 5(a) and (b)}
\end{table}

\begin{figure}[tbp]
   	\begin{center}
	\epsfig{figure=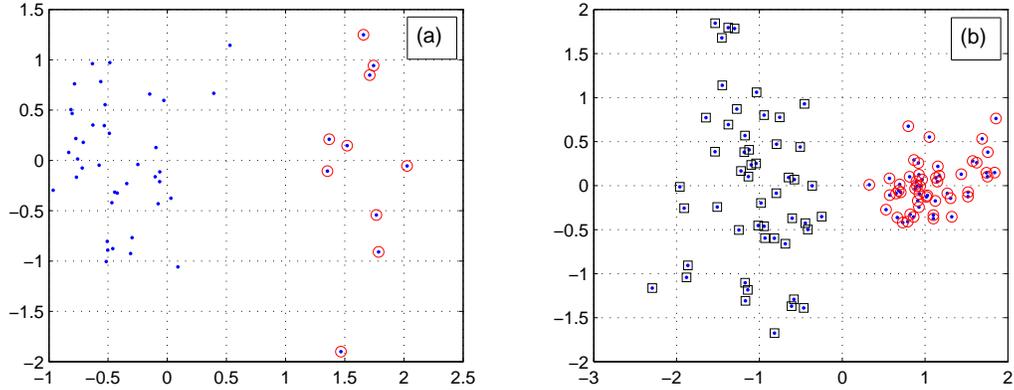,width=1\linewidth}
	\end{center}
     \caption{Distributions of mapped connection patterns of nodes in $2D$ feature space for two particular cases. The network structure is described as in text. In both cases, the (blue) dots represents all nodes of the network. In case (a), dots with (red) circles are corresponding to the nodes in densely connected group. In case (b), dots with (black) squares and (red) circles are corresponding to the nodes in two different large cluster ($50$ nodes each).} 
	\label{fig:fig6}
\end{figure}

The manifested prominent structures of a network depends on the investigation scale. A micro-strucrture of a cluster can be the dominant one at the appropriate scale. The proposed multi-phase method can adapt automatically according to the different scales.  An illustration is given in figure \ref{fig:fig6}. A densely connected cluster of size $M=10$ and average degree $\langle n\rangle=8$ is embedded in a random network of size $N=50$ and average degree $5$. The distribution of the projected connection patterns in feature space is shown in figure 6(a), which clearly reveal the structure of the network. However, when a larger network is studied, which consists of above subnetwork and other coexisting strong clusters, the structure in the subnetwork would become less important micro-structure in a cluster, and not the dominant structure of the whole network. Consider a larger network of size $N=100$, which consists of the above subnetwork of size $M=50$ and other $50$ nodes which form a densely connected cluster with average degree about $<n>=14$. The two groups are then sparsely connected (on average, one edge per node is added to connect to other group). Obviousely, the dominant structure is that of two clusters with similar sizes in the whole network. This is correctly reflected in figure 6(b). The micro-structure in the first cluster has been supressed in feature space by the coexisting more dominant structure in the larger scale. This characteristic makes the proposed method valuable when applying to network with hierarchical structures. 

In previous sections, we describe a specific implementation of the multi-phase analysis method. However, the essential merits of the proposed method do not rely much on specific similarity measurement and techniques (such as PCA) used. If more information on network can be incorporated, better tools exist for analysis in each phase depending on the problem at hand. It is the analysis strategy, i.e., working on certain feature space instead of the original network, that makes the proposed method a more general way to analyze network structures.

In summary, we propose here a novel multi-phase approach to analyze the network structures. By focusing on the clustering structure in feature space, we are able to circumvent several difficulties caused by the diversity of micro-structures and different scales. The method has been tested on several model networks which have not been extensively discussed up to now. The demonstrated advantages show that it can be applied to networks with more general structures. We believe that there are many situations in practice where the proposed method may be applied effectively.  

\section{Acknowledgment}

This work is supported by Temasek Laboratories at National University of Singapore through the DSTA Project POD0613356.


\begin{thebibliography}{99}

\bibitem{Hartwell1999}
Hartwell L. H., Hopfield J. J., Leibler S., and Murray A. W., Nature 499, C47-C52 (1999).
\bibitem{Ravasz2002}
Ravasz E. Somera A. L., Mongru D. A., Oltvai Z. N., and Barab\'{a}si A. L., Science 297, 1551-1555 (2002).
\bibitem{Jolliffe2002}
I. T. Jolliffe, "Principal Component Analysis" (2nd ed.), Springer, NY, 2002.

\bibitem{NewmanGirvan2004}
M. E. J. Newman and M. Girvan, Phys. Rev. E 69, 026113 (2004).
\bibitem{Newman2004}
M. E. J. Newman, Eur Phys. J. B 38, 321-330 (2004).
\bibitem{Newman2006}
M. E. J. Newman, Phys. Rev. E 74, 036104 (2006).
\bibitem{Duch2005}
J. Duch and A. Arenas, Phys. Rev. E 72, 027104 (2005).
\bibitem{Wu2004}
F. Wu and B. A. Huberman, Eur. Phys. J B 38, 331-338 (2004).
\bibitem{Reichardt2004}
J. Reichardt and S. Bornholdt, Phys. Rev. Lett. 93, 218701 (2004).
\bibitem{Jacob2007}
Jocob Kogan, "Introduction to Clustering Large and High-Dimensional Data", Cambridge University Press, 2007. 
\bibitem{Fukunaga1990}
Keinosuke Fukunaga, "Introduction to Statistical Pattern Recognition", Academic Press, 1990.

\bibitem{Newman2001}
M. Newman, S. Strogatz, and D. Watts, Phys. Rev. E 64, 026118 (2001).

\end{thebibliography}
\end{document}